\documentstyle[12pt,worldsci]{article}
\begin{document}
\def\z2{\ifmmode Z_2\else $Z_2$\fi}
\def\ie{{\it i.e.},}
\def\eg{{\it e.g.},}
\def\etal{{\it et. al.}}
\def\to{\rightarrow}
\def\tcv{\ifmmode t\to cV\else $t\to cV$\fi}
\def\Re{{\cal R \mskip-4mu \lower.1ex \hbox{\it e}\,}}
\def\Im{{\cal I \mskip-5mu \lower.1ex \hbox{\it m}\,}}
\pagestyle{empty}
\setlength{\baselineskip}{2.6ex}

\title{{\bf INVERSE NEUTRINOLESS DOUBLE $\beta$-DECAY AT THE NLC?}}
\author{T.G.~RIZZO\\
\vspace{0.3cm}
{\em High Energy Physics Division, Argonne National Laboratory, Argonne, IL
60439, USA}}
\maketitle

\begin{center}
\parbox{13.0cm}
{\begin{center} ABSTRACT \end{center}
{\small\hspace*{0.3cm}
The NLC may allow us to search for the `inverse' process to
neutrinoless double-$\beta$ decay, $e^-e^-\to W^-W^-$, which probes the
existence of Majorana masses for neutrinos.
Expectations for the observation of this process in both the Standard Model
and its extension, the Left-Right Symmetric Model, are examined.}}
\end{center}

Within the framework of extended electroweak models(EEM) one of the most
attractive scenarios for explaining the apparently small size of neutrino
masses is to invoke the see-saw mechanism. This scheme naturally predicts that
neutrino masses are of the Majorana type and that new heavy isosinglet
leptons($N$),
which can mediate $\Delta L$=2 interactions, must exist. At low energies,
such models can be probed indirectly by looking for rare processes such as
neutrinoless double-$\beta$ decay. The lack of observation of these processes
implies additional constraints on model building.

High energy $e^-e^-$ collisions, a possible option at the NLC, may provide a
new window into the $\Delta L$=2 sector of these models via the process
{\cite {1,2}} $e^-e^- \to W^-W^-$, where the $W$ is either the conventional
gauge boson of the Standard Model(SM) or that of some EEM. By forcing the
reconstructed
final state $W$ pairs to have large invariant masses and by imposing missing
energy and rapidity cuts, the SM backgrounds for such a process can be
reduced to
a level substantially below 0.1 fb for a 1 TeV $e^-e^-$ collider{\cite {3}}.
However, if we try to calculate the cross-section($\sigma$) for this process
assuming that only a single $N$ is exchanged in the $t-$ and $u-$channels,
one is immediately faced with the prospect of unitarity violation which can be
cured only by the exchange of additional particles{\cite {4}}. There are
potentially two distinct approaches: ($i$)in the single-generation SM, where
$N$ is identified with $\nu^c$, the weak eigenstates $\nu$ and $\nu^c$ mix, by
an angle $\theta$, to
form the two Majorana mass eigenstates $N_{1,2}$ which are now {\it both}
exchanged in the $e^-e^- \to W^-W^-$ process. By noting the relationships
between the mass matrix entries, $\theta$, and the mass eigenvalues, we find
that the leading high $s$ term in the amplitude is indeed cancelled resulting
in the restoration of unitarity. However, as a consequence of this, $\sigma$
is very small, as shown in Fig.~1, due to the enormous mixing angle
suppression. (This suppression
occurs since $\sigma$ is now proportional to $\theta^4$, and we would suspect
that $\theta<O(10^{-2})$). Although we can imagine that in a
multi-generational model there may be sufficient parameter freedom to
conspire to overcome
some of this suppression, we can anticipate that within the SM this will be
somewhat difficult to achieve and that the value of
$\sigma$ will likely remain small. We note, however, that $N$'s may also make
their existence known indirectly by a slight degradation of the forward
peak arising
from the $t-$channel amplitude in $e^+e^- \to W^+W^-$ of order a few percent.
Unfortunately, this will tell us {\it {nothing}} about their Dirac $vs.$
Majorana nature. Of course, if $N$'s can be directly produced at a collider,
their decays will tell us whether they are Dirac or Majorana particles.

($ii$)A second scenario to cure the unitarity problem is the $s$-channel
exchange of a doubly-charged Higgs scalar($\Delta$) which couples to $e^-e^-$.
This possibility is phenomenologically excluded in the SM due both to
$\rho$-parameter and neutrino counting constraints but can be realized
naturally in some EEM such as
the Left-Right Symmetric Model(LRM){\cite {5}}. (We can, of course, introduce
a $\Delta$ that couples to isosinglet right-handed electrons without violating
these constraints in the SM context but then $\Delta$ will not couple to $WW$
and so will not cure the unitarity problem.) In the LRM case, the sum of the
asymptotic $\nu$ and $\nu^c$ contributions to the amplitude no longer
cancel and
a $\Delta$ exchange is required in order to restore unitarity. Of course, the
$W$'s in the final state are now to be identified with $W_R$'s, the
right-handed gauge bosons of the LRM. For purposes of the
analysis presented here we will assume that $W_R$'s have a mass of 480 GeV and
are thus kinematically accessible at a 1 TeV $e^-e^-$ collider. We will also
assume that $\kappa=g_R/g_L=0.9$, consistent with $SO(10)$ renormalization
group analyses{\cite {6}}, and that $N$ has a mass in the 100 GeV and above
range thus avoiding the $W_R$ mass bounds from $\mu$ decay as well as
Tevatron collider searches.
(Similar constraints arising from the $K_L-K_S$ mass difference can also be
avoided but will not be discussed here.) Unitarity for $\sigma$'s in various
channels is
maintained so long as $N$($\Delta$) has a mass less than about 2(10) TeV.
Generally the cross section for $e^-e^- \to W^-_RW^-_R$ can be quite large, as
shown in Figs.~2a and 2b, with very sizeable
event rates $>10^{4-5}$ for integrated luminosities in the 100 $fb^{-1}$
range. (For most values of the parameters, the $cos~\theta$ distribution of
the produced $W_R$'s is fairly flat implying that kinematic cuts will
not significantly
reduce these rates.) Figs.~2a and 2b show that for small $M_N$ the cross
section tends to zero
because the amplitude is proportional to the Majorana mass itself
since it is the source of the explicit
lepton number violation. For larger $M_N$, there is found to be a trade
off between
the proportionality to $M_N$ in the numerator of the amplitude and the $M_N^2$
factor appearing in $u-$ and $t-$channel propagators. Except near the narrow
$\Delta$
resonance, the production cross section is {\it {relatively}} insensitive to
$M_{\Delta}$ as shown explicitly in Fig.~2b. Clearly, if the $W_R$ pair final
state is kinematically accessible at the NLC and if the $N$ and $W_R$ masses
are at all comparable, then the  process $e^-e^- \to W^-_RW^-_R$ should be
observable with a significant rate. We remind the reader that $\sigma$ scales
as $\kappa^4$ so that results for other values of $\kappa$ can be easily
obtained from the figures.

If this LRM scenario is indeed realized, one can imagine that a small amount
of $W-W_R$ mixing (by an angle $\phi$) will naturally be present and thus
could result in
the feeding of the large $W^-_RW^-_R$ rate into other channels. $\phi$ is
essentially given by $\phi \simeq  f \kappa M_L^2/M_R^2$, with $M_L(M_R)$
being the SM $W$($W_R$) mass and, in the minimal model, $f<1$ is a ratio
of squares of vev's so that numerically $\phi \simeq 0.01$. The most
obvious channel to examine in this connection is to see whether this
non-zero mixing
can induce a significant SM $W^-W^-$ rate. Unfortunately, as shown in
Fig.~3a, unless the $W-W_R$ mixing angle is very large, the induced $W^-W^-$
cross section will remain quite small.

Mixing in the $\nu-\nu^c$ mass matrix, while not significantly contributing to
$W^-_RW^-_R$ production, could induce the mixed final state $W^-W^-_R$ even in
the absence of $W-W_R$ mixing. This process will proceed only via $t-$ and
$u-$channel exchanges when $\phi=0$ (since $\Delta$ does not couple to the
weak eigenstate $WW_R$ combination) but will still obey unitarity due to the
opposite helicity structures at the two vertices. The cross section
for this mixed final state is shown in Fig.~3b and is seen to be reasonably
small but perhaps still observable depending on the value of the $\nu-\nu^c$
mixing angle, $\theta$, introduced above. If {\it {both}} $\phi$ and $\theta$
are simultaneously non-zero then the possibility of feed down becomes quite
complicated and we refer the reader to the very detailed analysis presented
in Ref.~7. Clearly, if $W^-W^-$ production is indeed observed, all possible
channels must be examined in order to pin down the nature of the lepton number
violating interaction.

In scenario ($ii$), realized in the LRM example above, other new physics must
also be present. An example of this is the process $e^+e^- \to \mu^+\mu^-$
which can occur via $s-$channel $\Delta$ exchange{\cite {8}}. For a $\Delta$
of mass 2 TeV and an $e^+e^-$ center of mass energy of ${\sqrt {s}}= 1$ TeV,
one finds a cross section, $\sigma \simeq 620 (\kappa M_N/M_R)^4~fb$, which
yields a very large event rate with a flat angular distribution for integrated
luminosities in the 10-100 $fb^{-1}$ range. Such a process would be quite
difficult to miss especially if a $\Delta$ resonance exists with a mass less
than the NLC's center of mass energy.

In conclusion we have found that $e^-e^-$ collisions at a center of mass
energy near 1 TeV may provide a useful probe of the neutrino mass generating
mechanism, in both the SM as well as EEM's, via the process
$e^-e^- \to W^-W^-$. Other new physics signatures may also be present in the
$e^-e^-$ channel.

\bibliographystyle{unsrt}

\begin{thebibliography}{99}
\bibitem{1} This possibility was first discussed in T.G.~Rizzo,
{\em Phys.~Lett.} {\bf B116}, 23 (1982).
%
\bibitem{2} D.~London, G.~Belanger, and J.N.~Ng, {\em Phys.~Lett.} {\bf B188},
155 (1987; J.~Maalampi, A.~Pietil\" a, and J.~Vouri, {\em Phys.~Lett.}
{\bf B297}, 327 (1992) and Turku University report FL-R9 (1992); M.P.~Worah,
Enrico Fermi Institute report EFI 92-65 (1992);
D.A.~Dicus, D.~Dzialo, and P.~Roy, {\em Phys.~Rev.} {\bf D44}, 2033 (1991);
C.A.~Heusch and P.~Minkowski, CERN report CERN-TH-6606 (1992).
%
\bibitem{3} This estimate is based on the analysis of J.F.~Gunion and
A.~Tofighi-Niaki, {\em Phys.~Rev.} {\bf D36}, 2671 (1987) and {\bf D38}, 1433
(1988).
%
\bibitem{4} Generic cross section formulae are given in Ref.~1.
%
\bibitem{5} For a review and original references, see R.N.~Mohapatra,
{\it {Unification and Supersymmetry}}, (Springer, New York, 1986).
%
\bibitem{6} N.G.~Deshpande, E.~Keith, and T.G.~Rizzo, {\em Phys.~Rev.~Lett.}
{\bf 70}, 3189 (1993).
%
\bibitem{7} For general formulae, see J.~Maalampi $etal.$, Ref.~2.
%
\bibitem{8} See, for example, T.G.~Rizzo, {\em Phys.~Rev.} {\bf D25}, 1355
(1982).
%

\end{thebibliography}

{\small
\vspace*{2.25in}
\noindent
Fig.~1: Total cross section for $e^-e^- \to W^-W^-$ in the SM as a function
of the heavy neutrino mass with ${\sqrt {s}}$= 500(solid) or 1000(dash-dot)
GeV. Results should be scaled by 4 powers of the mixing angle $\theta$.

\vspace*{2.25in}
\noindent
Fig.~2: Cross section for $e^-e^- \to W^-_RW^-_R$ with ${\sqrt {s}}$=1 TeV
as a function of (a)$M_N$
and (b)$M_{\Delta}$ for the parameter choices discussed in the text.
In[(a),(b)],
the curves on the right(left)-hand side correspond, from top to bottom, to
$M_{\Delta}$=800, 1200, 500, 1500, 200, and 2000 GeV [$M_N$= 1500, 1200, 800,
500, 200 GeV].

\vspace*{2.25in}
\noindent
Fig.~3: (a)$W^-W^-$ production via $W-W_R$ mixing as a function of $M_N$ with
$M_{\Delta}$=300, 150, 800, 1200, 1500, 2000 GeV for the curves from top to
bottom on the right-hand side assuming ${\sqrt {s}}$=500 GeV. (b) $W^-W^-_R$
production as a function of $M_N$ with ${\sqrt {s}}$=1 TeV. This result must
be rescaled by 2 powers of the mixing angle ratio $\theta/0.01$.
 }

\end{document}